\documentclass[showpacs,superscriptaddress,nofootinbib]{revtex4}

\usepackage{graphicx}
\usepackage{graphics}
\usepackage{xspace}
\usepackage{amssymb}
\usepackage{amsmath}
\usepackage{latexsym}
\usepackage{mathrsfs}
\usepackage{pstricks}

\newcommand{\ii}{{\rm i}}
\newcommand{\de}{{\rm\,d}}
\newcommand{\e}{{\rm e}}

\newcommand{\tr}{\mbox{Tr}\,}
                 
\newcommand{\st}{{\scriptscriptstyle T}}

\newcommand{\g}{\gamma}

\newcommand{\Sslash}{\rlap{/} S}

\newcommand{\kslash}{\rlap{/} k}

\newcommand{\sca}{{s}}
\newcommand{\ps}{{v}}

\begin{document}
 

\title{
Sivers function in a spectator model with axial-vector diquarks}

\author{Alessandro Bacchetta}
\email{alessandro.bacchetta@physik.uni-regensburg.de}
\affiliation{Institut f\"ur Theoretische Physik, Universit\"at Regensburg,
D-93040 Regensburg, Germany}

\author{Andreas Sch\"afer}
\email{andreas.schaefer@physik.uni-regensburg.de}
\affiliation{Institut f\"ur Theoretische Physik, Universit\"at Regensburg,
D-93040 Regensburg, Germany}

\author{Jian-Jun Yang}
\email{jian-jun.yang@physik.uni-regensburg.de}
\affiliation{Institut f\"ur Theoretische Physik, Universit\"at Regensburg,
D-93040 Regensburg, Germany}

\date{\today}

\begin{abstract}
We perform a calculation of the Sivers function in a spectator model of the
nucleon, with scalar and axial-vector diquarks. 
We make use of gluon rescattering 
to produce the nontrivial phases necessary to generate the Sivers function. 
The inclusion of axial-vector diquarks 
enables us to obtain a nonzero  
Sivers 
function for down quarks. 
Using the results of our model, 
we discuss the phenomenology of transvere single spin asymmetries
in $\pi^+$, $\pi^-$, and $\pi^0$ production,
which are 
currently analysed
by the HERMES and COMPASS collaborations. 
We find that the inclusion of axial-vector diquarks substantially reduces the
asymmetries. 
\end{abstract}


\maketitle

\section{Introduction}

The Sivers function was introduced for the first time in Ref.~\cite{Sivers:1990cc}, 
in an attempt to explain the observation of single-spin asymmetries in hard
hadronic reactions. Since then, some phenomenological extractions of the
Sivers function have been
performed~\cite{Anselmino:1998yz,Anselmino:1999pw,Anselmino:2002pd}, from data 
on pion production in proton-proton
collisions~\cite{Adams:1991cs,Bravar:1996ki}.
The Sivers function gives a contribution also to the single-spin asymmetry
observed by the HERMES collaboration 
in pion production via deep inelastic scattering off polarized
targets~\cite{Airapetian:1999tv,Airapetian:2001eg,Airapetian:2002mf}. In all
the above cases, however, 
the presence of competing
effects (in particular the Collins effect) did not allow clear
conclusions up to now~\cite{Anselmino:1999pw,Efremov:2003tf}. 
 
Despite the phenomenological indications, for several years the Sivers
function was believed to vanish due to time-reversal
invariance~\cite{Collins:1993kk}. However, 
this statement was contradicted by an explicit calculation by Brodsky,
Hwang and Schmidt, using a
spectator model~\cite{Brodsky:2002cx}.  
As the Sivers function is an example of T-odd entity, it
requires the interference 
between two amplitudes with different imaginary
parts~\cite{Collins:1993kk,Bacchetta:2001di}.  
 Spectator models at
tree level cannot provide these nontrivial phases, but they can 
arise as soon as
a gluon is exchanged between the struck quark and the target
spectator~\cite{Brodsky:2002cx}. 
More generally, the presence of the gauge link, which
insures the color gauge 
invariance of parton distributions, can provide nontrivial phases and thus 
generate T-odd
functions~\cite{Collins:2002kn,Ji:2002aa,Belitsky:2002sm,Boer:2003cm}. The
main ingredient of the model calculation of \cite{Brodsky:2002cx} 
is nothing else than 
the one-gluon approximation to the gauge link.
It is also interesting to note that 
T-odd distribution functions vanish in a large class of chiral soliton models, 
where
 gluonic degrees of freedom are absent~\cite{Pobylitsa:2002fr}.

The work of Brodsky, Hwang and Schmidt was not aimed at producing a
phenomenological estimate. A step forward in this direction has been
accomplished in Refs.~\cite{Boer:2002ju,Gamberg:2003ey}.
In our article, we present an alternative
calculation, using the version of the spectator model presented by
Jakob, Mulders and Rodrigues in Ref.~\cite{Jakob:1997wg}. In particular, we
include in the model a dynamical 
axial-vector diquark as a possible spectator, and we
explore the Sivers function for down
quarks.  The necessity to include axial-vector diquarks is also discussed,
e.\ g., in
Refs.~\cite{Close:1988br,Meyer:1991fr,Oettel:2000jj}.
Finally, we point out that a calculation of the Sivers function in the MIT bag
model has been recently presented
in Ref.~\cite{Yuan:2003wk}.

\section{Unpolarized distribution function $f_1$}

The unpolarized distribution function $f_1$ can be defined as
\begin{equation}
f_1 (x,\vec{k}_T^2) =\frac{1}{4}\, \tr\bigl[\bigl(\Phi(x,\vec{k}_T; S)+\Phi(x,\vec{k}_T; -S)\bigr)\,\g^+\bigr]\,,
\end{equation} 
where $S$ is the spin of the target.
The correlator $\Phi(x,\vec{k}_T)$ can be written as~\cite{Boer:2003cm}
\begin{equation}  
\Phi(x,\vec{k}_\st; S)=\int
        \frac{\de\xi^- \de^2\xi_\st}{(2\pi)^{3}}\;
 \e^{+\ii k \cdot \xi}
       \langle P, S|\bar{\psi}(0)\,{\cal L}_{[0^-, \infty^-]}
{\cal L}_{[0_\st, \infty_\st]}{\cal L}_{[\infty_\st, \xi_\st]}{\cal
L}_{[\infty^-, \xi^-]}\psi(\xi)|P, S \rangle \bigg|_{\xi^+=0}\,,
\label{e:phi}
 \end{equation}  
where the notation ${\cal L}_{[a,b]}$ indicates a straight gauge link running 
from $a$ to $b$. In Drell-Yan processes the link runs in the opposite
direction, to $-\infty$~\cite{Boer:2003cm}. 
For the calculation of the unpolarized function $f_1$ the transverse part of
the gauge link does not play a role and the entire gauge link can be reduced
to unity. Therefore, for this first part of the calculation it is sufficient
to consider only
the handbag diagram.

\begin{figure}
	\centering
        \includegraphics[width=10 cm]{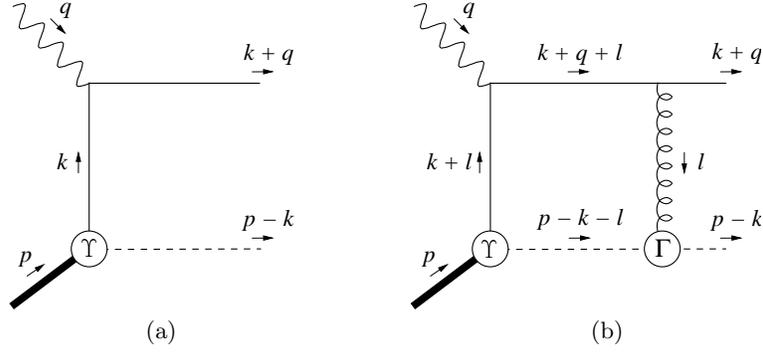}\\
	(a)\hspace{55mm}(b)	
        \caption{Tree-level and one-loop diagrams for the specator-model 
		calculation of the Sivers function. The dashed line indicates
		both the scalar and axial-vector diquarks.}
        \label{f:diagsivers}
        \end{figure}

At tree level, we follow almost exactly the spectator model of Jakob, Mulders
and Rodrigues~\cite{Jakob:1997wg}. In this model, the proton (with mass $M$)
can couple to a
constituent quark of mass $m$ and a diquark. The diquark can
be both a scalar particle, with mass $M_\sca$, or an axial-vector particle, 
with mass $M_\ps$. The relevant diagram at tree level (identical for the
scalar and axial-vector case) is depicted in
Fig.~\ref{f:diagsivers} (a). 
In our model, the nucleon-quark-diquark vertices are
\begin{align}
\Upsilon_\sca & = g_\sca(k^2), & \Upsilon_{\ps}^{\mu} & = \frac{g_\ps
(k^2)}{\sqrt{2}}\,\g_5 \g^{\mu}.
\end{align} 
We make use of the dipole form factor
\begin{equation}
 g_{\sca/\ps}(k^2) = N_{\sca/\ps} \frac{(k^2 -
m^2)\,(1-x)^2}{\bigl(\vec{k}_T^2 + L_{\sca/\ps}^2\bigr)^2},
\label{e:formfac}
\end{equation} 
where
\begin{align}
\vec{k}_T^2&=- (1-x)\, k^2 -x\, M_{\sca/\ps}^2 +x \,(1-x)\, M^2, \\
L_{\sca/\ps}^2&=  (1-x)\,\Lambda^2 + x\, M_{\sca/\ps}^2 -x\, (1-x)\, M^2. 
\end{align} 
The only difference with respect to Ref.~\cite{Jakob:1997wg} is the form of
$\Upsilon_{\ps}$ -- the 
vertex involving nucleon, quark, and axial-vector diquark. This change
modifies the original results only slightly.
Note that our choice of the form factor, defined in Eq.~(\ref{e:formfac}), is
very different from the Gaussian form factor employed in
Ref.~\cite{Gamberg:2003ey}. Both choices have the effect of eliminating the
logarithmic divergences arising from $k_T$ integration and suppress the
influence of 
the high $k_T$ region, where anyway
perturbative corrections should be taken into account~\cite{Boer:2002ju}.

The final results for the unpolarized distribution function $f_1$ are
\begin{align}
\begin{split}
f_1^\sca (x, \vec{k}_T^2)&= \frac{g_\sca^2 \,\bigl[(x M +m)^2 + \vec{k}_T^2
\bigr]}{2\, (2 \pi)^3\,(1-x)\, (k^2-m^2)^2} 
\;
= \frac{N_\sca^2\, (1-x)^3\,\bigl[(x M +m)^2 + \vec{k}_T^2\bigr]}{16 \pi^3\,\bigl(\vec{k}_T^2 + L_\sca^2\bigr)^4}\,,
\end{split} \\
\begin{split}
f_1^\ps (x, \vec{k}_T^2)&= \frac{g_\ps^2\,\bigl[(x M +m)^2 + \vec{k}_T^2 + 2xmM
\bigr] }{2\, (2 \pi)^3\, (1-x)\,(k^2-m^2)^2 } 
\;
= \frac{N_\ps^2\, (1-x)^3\,\bigl[(x M +m)^2 + \vec{k}_T^2+ 2xmM\bigr]}{16 \pi^3\,\bigl(\vec{k}_T^2 + L_\ps^2\bigr)^4}.
\end{split}
\end{align} 
Both functions can be integrated over the transverse momentum to give
\begin{align} 
f_1^\sca (x) & = \frac{N_\sca^2\, (1-x)^3}{96\,
\pi^2\, L_\sca^6} \,\bigl[2(x M +m)^2 + L_\sca^2\bigr], \\
f_1^\ps (x) & = \frac{N_\ps^2\, (1-x)^3}{96\,
\pi^2\, L_\ps^6} \,\bigl[2(x M +m)^2 + L_\sca^2 + 4xmM \bigr].
\end{align} 

In order to obtain the distribution functions for $u$ and $d$ quarks, we use
the following relation, coming from the analysis of the proton wave function,
\begin{align}
f_1^u & = \frac{3}{2}\, f_1^\sca + \frac{1}{2}\, f_1^\ps, &
f_1^d & = f_1^\ps.
\label{e:ud}
\end{align} 

\begin{figure}
	\centering
        \includegraphics[width=10 cm]{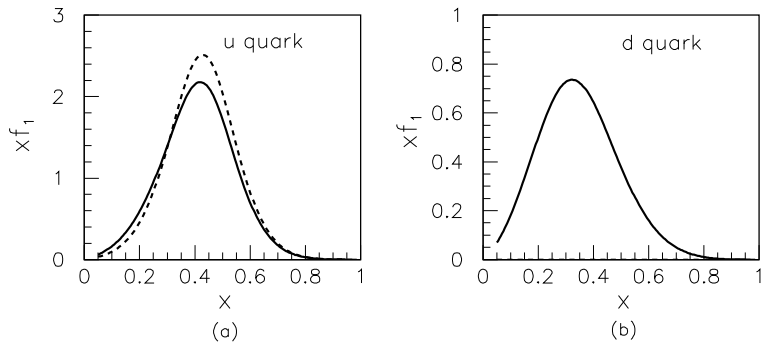}
        \caption{Model calculation of
		$x f_1 (x)$:
		with scalar diquarks only (dashed
		line),  with scalar and axial-vector diquarks (solid
		line). The $d$ quark distribution is zero when only 
		scalar diquarks are used.}
        \label{f:f1}
        \end{figure}

Here we refrain from discussing the choice of parameters of the model and its
quality, for which we refer to the original work~\cite{Jakob:1997wg}. 
We choose the following values for the parameters of the model:
\begin{align}
m&= 0.36\; {\rm GeV},& M_\sca&=0.6\; {\rm GeV}, & M_\ps&=0.8\; {\rm GeV}, \\
\Lambda&=0.5\; {\rm GeV}, & N_\sca^2&= 6.525, & N_\ps^2&= 28.716\, .
\end{align}
The factors $N_\sca$ and $N_\ps$ are chosen in order to
normalize the functions  $f_1^\sca$ and $f_1^\ps$  to 1 and consequently to
normalize $f_1^u$ to 2 and $f_1^d$ to 1.
The results of the model are shown in Fig.~\ref{f:f1}. 
The dashed line
represents the result of the spectator model with scalar diquarks only (with
$f_1^u = 2 f_1^\sca$). As
can be seen, the difference for the $u$ distribution is not big, but it is
particularly relevant at small $x$. The $d$ quark distribution is zero when
only scalar diquarks are used, which is clearly unrealistic.

\begin{figure}
	\centering
        \includegraphics[width=10 cm]{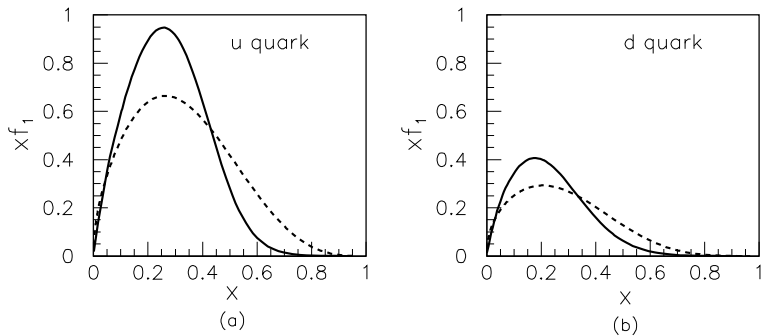}
        \caption{Model calculation of
		$x f_1 (x)$ (solid line) compared to the CTEQ5L
		parametrization~\cite{Lai:1999wy}  (dashed line) at 1 GeV$^2$.}
        \label{f:f1p}
        \end{figure}

One of the problems when trying to match the
model and the phenomenology is that it is not clear at which energy scale 
the model should be applied. 
A way to estimate this energy scale is to 
compare the total momentum carried by the valence quarks in the model
and in some parametrization~\cite{Meyer:1991fr,Bacchetta:2000dc}. 
Taking, for instance, the CTEQ5L
parametrization~\cite{Lai:1999wy}\footnote{We use leading
order evolution with three flavors and $\Lambda_{LO}^{(3)} = 0.222$, 
in order to match the CTEQ5 results.}
it turns out that this scale is about 0.078 GeV$^2$.
Then, by applying standard DGLAP equations, we can 
evolve our model results to 1 GeV$^2$ and compare it with
the CTEQ5L parametrization at that scale. The result is
shown in Fig.~\ref{f:f1p}.
Admittedly, the model reproduces the parametrization of the valence quark
distribution 
only qualitatively. 
In any case, in this work we mainly aim at giving rough estimates of
the relative magnitude of the $u$ and $d$ 
Sivers function and of the related single-spin asymmetries, as well as
studying the changes induced 
in the model results when an axial-vector diquark is introduced. Therefore, we
refrain from improving and tuning the model.

\section{Sivers function and its moments}

We use the following definition of the Sivers function
\begin{equation}
\frac{\epsilon^{ij}_T k_{Ti} S_{Tj}}{M}\, f_{1T}^{\perp} (x,\vec{k}_T^2) = - \frac{1}{4}\,
\tr\bigl[\bigl(\Phi(x,\vec{k}_T; S)-\Phi(x,\vec{k}_T; -S)\bigr)\,\g^+\bigr]\,,
\end{equation} 
where $ 4 \ii\, \epsilon^{ij}_T k_{Ti} S_{Tj} = \tr[\g_5 \g^+ \g^- \kslash\,
\Sslash]$. 
At tree level the Sivers function turns out to vanish. This is due to the lack 
of any final state interaction that can provide the imaginary parts necessary
to generate T-odd functions. We need to 
introduce the one-loop amplitude described
in Fig.~\ref{f:diagsivers} (b). This is nothing else than the one-gluon
approximation to the gauge link included in Eq.~(\ref{e:phi}). The Sivers
function receives a contribution from the interference between amplitude (a)
and the imaginary part of amplitude (b). The
imaginary part of amplitude (b) can be computed by applying Cutkosky rules
 or,
equivalently, by taking the imaginary part of the propagator $1/(l^+ + \ii
\epsilon)$~\cite{Ji:2002aa}. Note that in Drell-Yan processes the different
topology of the one-gluon diagram implies that the imaginary part of the
propagator  $1/(l^+ - \ii
\epsilon)$ has to be taken, with the effect of changing the overall sign of
the Sivers function~\cite{Collins:2002kn,Brodsky:2002rv}. This is consistent with the change in direction of the
gauge link mentioned before~\cite{Boer:2003cm}.

Following Ref.~\cite{Brodsky:2002cx} we perform the calculations initially 
with Abelian gluons and generalize the result to QCD at the end. We use
Feynman gauge.
In order to compute the one-loop diagram, we have to make the appropriate
choice for the vertex between the gluon and the scalar or axial-vector
diquark.
We choose the following forms
\begin{align}
\Gamma_\sca^{\mu} &= -\ii e_2 \bigl(2p -2k -l\bigr)^{\mu}, \\
\Gamma_\ps^{\mu, \alpha \beta} &=\ii e_2 \bigl[\bigl(2p -2k -l\bigr)^{\nu} g^{\alpha \beta} -
\bigl(p-k-vl\bigr)^{\beta} g^{\nu \alpha}- \bigl(p-k-(1-v)l\bigr)^{\alpha} g^{\nu \beta}\bigr],
\end{align} 
where $e_2$ denotes the color charge of the diquark, which we assume to be 
the
same for both kinds of diquark.
The gluon-axial-vector diquark coupling is identical to the
photon-axial-vector diquark coupling suggested in
Ref.~\cite{Oettel:2000jj}. The parameter $v$ is the anomalous magnetic
moment of the axial-vector diquark; for $v=1$ the vertex is analogous, for
instance, to the
standard photon-$W$ vertex. In any case, our results at leading order do 
not depend on the anomalous magnetic moment of the diquark.

The final results for the Sivers function are
\begin{align}
f_{1T}^{\perp \sca}(x,\vec{k}^2_T)&= \frac{e_1 e_2\,N_\sca^2\,(1-x)^3 \,M\,(xM+m)}
	{4\, (2 \pi)^4\,L_\sca^2\,\bigl[\vec{k}_T^2+ L_\sca^2 \bigr]^3},\\
f_{1T}^{\perp \ps}(x,\vec{k}^2_T)&= -\frac{e_1 e_2\,N_\ps^2\,(1-x)^3 \,xM^2}
	{8\, (2 \pi)^4\,L_\ps^2\,\bigl[\vec{k}_T^2+ L_\ps^2 \bigr]^3}.
\end{align} 
Note that our result
for the scalar diquark has the opposite sign compared to similar
computations~\cite{Brodsky:2002cx,Boer:2002ju}. 
However, a sign error in those computations has been
recently identified~\cite{Burkardt:2003je}. The other differences between our
results and those in Refs.~\cite{Brodsky:2002cx,Boer:2002ju,Gamberg:2003ey}
are due to the different choice of form factors in the nucleon-quark-diquark
vertex. 
Unfortunately, we cannot evolve our results to a higher energy scale,
as we have done for the unpolarized distribution function, since the
evolution equations for the Sivers function have not been established
yet~\cite{Henneman:2001ev,Kundu:2001pk}. We hope, however, that the $Q^2$
dependence of our results is not very strong.  

We introduce the ${k}_T$ moments
\begin{align} 
f_{1T}^{\perp (1/2)}(x) &\equiv \int \de^2 \vec{k}_T\; \frac{\bigl\lvert\vec{k}_T\bigr\rvert}{2M}\;f_{1T}^{\perp}(x,\vec{k}^2_T),&
f_{1T}^{\perp (1)}(x) &\equiv \int \de^2 \vec{k}_T \;\frac{\vec{k}_T^2}{2M^2}\; 
f_{1T}^{\perp}(x,\vec{k}^2_T),
\end{align} 
for which we get the following results
\begin{align} 
f_{1T}^{\perp (1/2) \sca}(x) &=  \frac{e_1 e_2\,N_\sca^2\,(1-x)^3 \,(xM+m)}
	{1024 \,\pi^2\,L_\sca^5} ,
&
f_{1T}^{\perp (1) \sca}(x) &=  \frac{e_1 e_2\,N_\sca^2\,(1-x)^3 \,(xM+m)}
	{256 \,\pi^3\,M\,L_\sca^4} ,
\\
f_{1T}^{\perp (1/2) \ps}(x) &=  -\frac{e_1 e_2\,N_\ps^2\,(1-x)^3 \,xM}
	{2048 \,\pi^2\,L_\ps^5} ,
&
f_{1T}^{\perp (1) \ps}(x) &=  -\frac{e_1 e_2\,N_\ps^2\,(1-x)^3 \,x}
	{512 \,\pi^3\,L_\ps^4} .
\end{align} 
The only parameter to be fixed is the product of the quark and diquark
charges. Following  Ref.~\cite{Brodsky:2002cx} we fix $e_1 e_2 = 4 \pi\, C_F\,
\alpha_s$ and we choose $C_F = 4/3$ and $\alpha_S \approx 0.3$.

\begin{figure}
	\centering
        \includegraphics[width=10 cm]{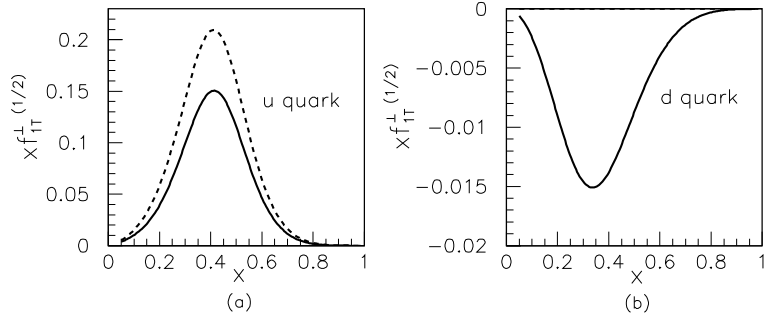}
        \caption{Model calculation of 
		$x f_{1T}^{\perp (1/2)}(x)$:
		with scalar diquarks only (dashed
		line),  with scalar and axial-vector diquarks (solid
		line). The $d$ quark distribution is zero when only 
		scalar diquarks are used.}
        \label{f:f1tperp}
        \end{figure}

Relations equivalent to those of Eq.~(\ref{e:ud}) hold for the Sivers function and 
its moments. In Fig.~\ref{f:f1tperp} we show the model results for the first
moment of the Sivers function. The inclusion of the axial-vector diquark
results in some change to the $u$ Sivers function and allows us
to produce a nonzero $d$ quark Sivers function. It turns out in
particular that the $d$ Sivers function is much smaller than the $u$
one and has the opposite sign. This result is in qualitative agreement with the
bag-model calculation of Ref.~\cite{Yuan:2003wk}.
The opposite sign of the two 
functions is also
in agreement with the only phenomenological extractions available at
present~\cite{Anselmino:1998yz,Anselmino:1999pw,Boglione:1999pz}. However,
there is a sharp difference in the 
relative magnitude of the two contributions, since in the phenomenological
extractions the absolute value of the $d$ contribution is about half of
the $u$ contribution, while in our model calculation the $d$ contribution is
only about $1/10$ of 
the $u$ contribution. We point out that this difference could be due to a
sizeable contribution of sea quarks (in particular $\bar{u}$) to the
asymmetry studied in Refs.~\cite{Anselmino:1998yz,Anselmino:1999pw}.

Note that our model calculation complies with the positivity bound~\cite{Bacchetta:2000kz}
\begin{equation}
f_{1T}^{\perp (1/2)}(x)\leq \frac{1}{2} f_1 (x)\, .
\label{e:boundint}
\end{equation}

We performed also the calculation of the function $h_1^{\perp}$, introduced by 
Boer and Mulders in Ref.~\cite{Boer:1998nt} and defined as
\begin{equation}
\frac{\epsilon^{ij}_T k_{Tj}}{M}\,
h_1^{\perp} (x,\vec{k}_T^2) =\frac{1}{4}\, \tr\bigl[\bigl(\Phi(x,\vec{k}_T;
S)+\Phi(x,\vec{k}_T; -S)\bigr)\,\ii \, \sigma^{i +} \g_5  \bigr]\,.
\end{equation} 
For the scalar diquark, we found that $h_1^{\perp \sca}=f_{1T}^{\perp \sca}$, confirming
the results already obtained in Refs.~\cite{Gamberg:2003ey,Boer:2002ju}.
For the axial-vector diquark, we obtain
\begin{align} 
h_1^{\perp \ps}(x,\vec{k}^2_T)&= - \frac{e_1 e_2\,N_\ps^2\,(1-x)^3 \,M\,(2 xM+m)}
	{4\, (2 \pi)^4\,L_\ps^2\,\bigl[\vec{k}_T^2+ L_\ps^2 \bigr]^3},
&
h_1^{\perp (1) \ps}(x) &=  -\frac{e_1 e_2\,N_\ps^2\,(1-x)^3 \,(2 xM+m)}
	{256 \,\pi^3\,M\,L_\ps^4}.
\end{align} 

\section{Single spin asymmetries}

\begin{figure}
	\centering
	\includegraphics[width=9cm]{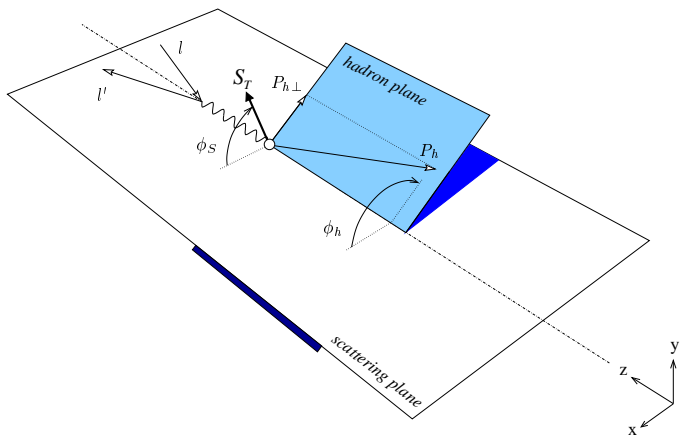}
	\caption{Description of the vectors and angles involved in the Sivers
		asymmetry measurement.}
	\label{f:angles}
        \end{figure}

 We consider the 
 weighted transverse spin asymmetries
\begin{align} 
\left\langle \sin{(\phi_h - \phi_S)} \right\rangle_{UT} (x,z) 
&= 	\frac{\int \de y\, \de \phi_S\, \de^2 \vec{P}_{h\perp}\,
	\sin{(\phi_h - \phi_S)}\; (\de^6\sigma_{U\uparrow} - \de^6\sigma_{U\downarrow})}
	{\int \de y\,  \de \phi_S\, \de^2 \vec{P}_{h\perp}
	(\de^6\sigma_{U\uparrow} + \de^6\sigma_{U\downarrow})}\, , 
\\
\Big\langle \frac{|\vec{P}_{h\perp}|}{z M}\, \sin{(\phi_h - \phi_S)}
\Big\rangle_{UT} (x,z)
&= \frac{\int \de y\,  \de \phi_S \de^2 \vec{P}_{h\perp}
\,|\vec{P}_{h\perp}|/(z M)\,\sin{(\phi_h - \phi_S)}\;
(\de^6\sigma_{U\uparrow} - \de^6\sigma_{U\downarrow})}
{\int  \de y\, \de \phi_S \de^2 \vec{P}_{h\perp}
(\de^6\sigma_{U\uparrow} + \de^6\sigma_{U\downarrow})},
\label{e:wasymm}
\end{align} 
where the notation $\de \sigma_{U\uparrow}$ indicates the cross section
with an unpolarized lepton beam off a transversely polarized
target. The angles involved in the definition of the asymmetry are depicted in 
Fig.~\ref{f:angles}.
These asymmetries are currently  measured by the HERMES and COMPASS
collaborations~\cite{Schnell:2003,Pagano:2003}. 

\begin{figure}
	\centering
        \includegraphics[width=10 cm]{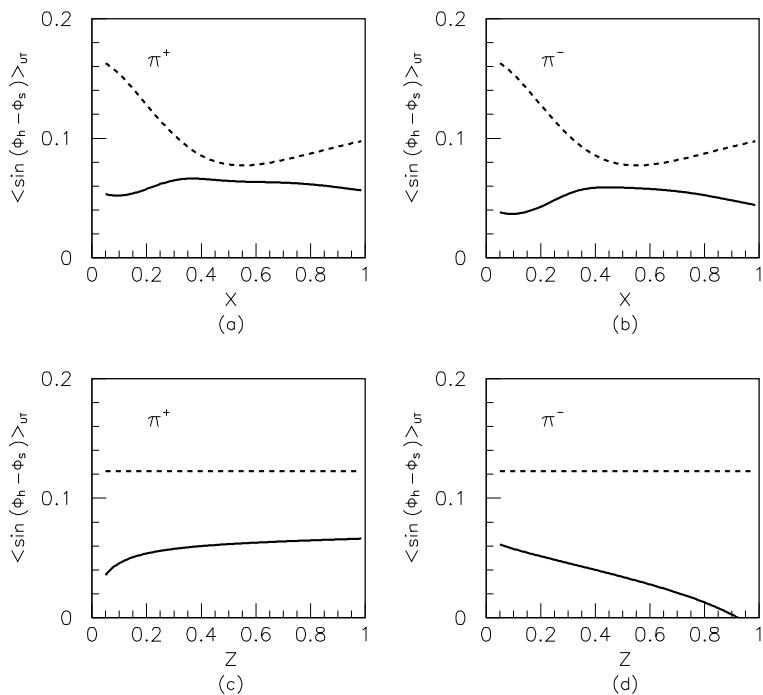}
        \caption{Model estimate of the 
	single-spin asymmetry $\langle\sin{(\phi_h-\phi_S)}\rangle_{UT}$: 
		with scalar diquarks only (dashed
		line),  with scalar and axial-vector diquarks (solid line). 
		The $x$ and $z$ dependence is shown
		for $\pi^+$ and $\pi^-$. }
        \label{f:asymm}
        \end{figure}

Under the assumption that the pion transverse momentum with respect to
the virtual photon is 
entirely due to the intrinsic transverse momentum of partons, i.e., 
$\vec{P}_{h\perp}= z \vec{k}_T$, the first asymmetry can be written as
\begin{equation} 
\left\langle \sin{(\phi_h - \phi_S)} \right\rangle_{UT} (x,z) \approx
\frac{1/x \,\sum_a e_a^2\, f_{1T}^{\perp (1/2)a}(x)\,D_1^a(z)}
{1/x \,\sum_a e_a^2\, f_1^a
(x)\,D_1^a(z)}\,,
\end{equation} 
with $a$ indicating the quark flavor.
Our model results are displayed in Fig.~\ref{f:asymm}. We took the unpolarized 
fragmentation functions from Ref.~\cite{Kretzer:2001pz} at a scale $Q^2=1 \,
\rm{GeV}^2$.  In order to make
predictions useful for the  
HERMES experiment, to produce
Fig.~\ref{f:asymm} (a) and (b) we integrated the asymmetries over $z$ from 0.2 to
0.7. To produce
Fig.~\ref{f:asymm} (c) and (d) we integrated the asymmetries over $x$ from 0.023 to
0.4. 

Evidently, there is not a big difference between $\pi^+$ and $\pi^-$
asymmetries. 
We do not plot the results for $\pi^0$ production, since they lie 
between the previous two.
We point out that assuming 
\begin{equation} 
f_1^d D_1^{d(\pi^{\pm,0})} \ll 4 \,f_1^u D_1^{u(\pi^{\pm,0})},
\label{e:approx}
\end{equation} 
the above asymmetry can be written as
\begin{equation} 
\left\langle \sin{(\phi_h - \phi_S)} \right\rangle_{UT}^{\pi^{\pm,0}} (x,z) \approx
\frac{3}{2}\,\frac{f_{1T}^{\perp (1/2)\sca}(x)}{f_1^u(x)}
+ \frac{1}{2}\,\frac{f_{1T}^{\perp (1/2)\ps}(x)}{f_1^u(x)}\,
\biggl(1 +
\frac{1}{2}\,\frac{D_1^{d(\pi^{\pm,0})}(z)}{D_1^{u(\pi^{\pm,0})}(z)}\biggr).
\label{e:asysimple}
\end{equation} 
The first term in
this equation is the dominant one. This explains
why there are only small differences
between the $\pi^+$ and $\pi^-$ asymmetry. However, the axial-vector contribution
to $f_1^u$  cannot be neglected.
From Eq.~(\ref{e:asysimple}) it is also evident that 
the dependence of the asymmetries on $z$ is due to the 
influence of the
unfavoured fragmentation functions. In the case of $\pi^-$ the last term in 
Eq.~(\ref{e:asysimple}) is
bigger and therefore the $z$ dependence is stronger. In the case of $\pi^0$ -- 
not shown in our pictures -- the term in parentheses 
would be exactly $3/2$ and the
$z$-dependent asymmetry would be a flat line at about $5\%$. 

\begin{figure}
	\centering
        \includegraphics[width=10 cm]{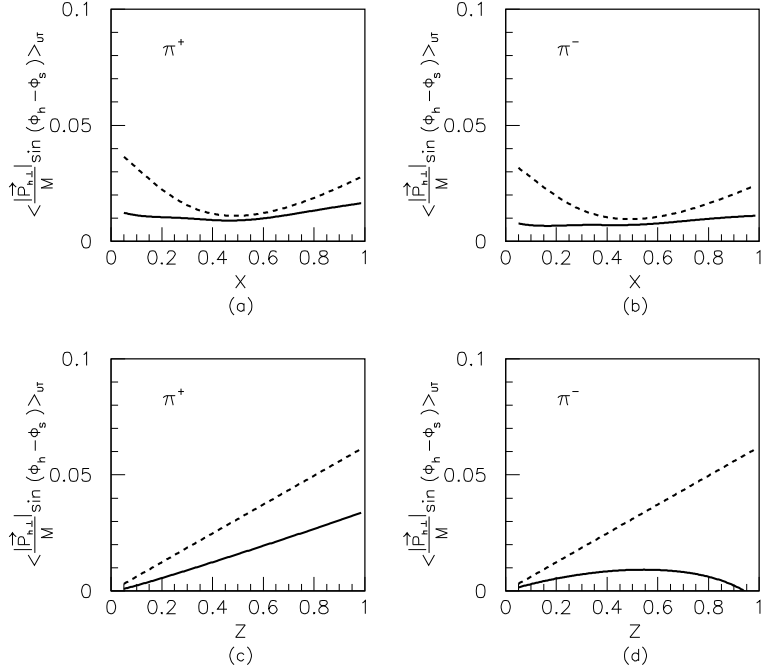}
        \caption{Model estimate of the 
		single-spin asymmetry 
		$\langle \frac{|P_{h \perp}|}{M}
		\,\sin{(\phi_h-\phi_S)}\rangle_{UT}$:
		with scalar diquarks only (dashed
		line),  with scalar and axial-vector diquarks (solid line). 
		The $x$ and $z$ dependence is shown
		for $\pi^+$ and $\pi^-$. }
        \label{f:wasymm}
        \end{figure}

The asymmetry in Eq.~(\ref{e:wasymm}) can be written in an assumption-free way
as
\begin{equation} 
\Bigl\langle \frac{|\vec{P}_{h\perp}|}{M}\, \sin{(\phi_h - \phi_S)}
\Bigr\rangle_{UT} (x,z)
= 
\frac{1/x \,\sum_a e_a^2\, f_{1T}^{\perp (1) a}(x)\,z\,D_1^a(z)}
{1/x \,\sum_a e_a^2\, f_1^a
(x)\,D_1^a(z)}\,.
\end{equation} 
The calculated asymmetries are shown in Fig.~\ref{f:wasymm}, where we
performed the integrations
over $x$ or $z$ as in the previous case.
As before, assuming Eq.~(\ref{e:approx}) the asymmetry can be simplified to
\begin{equation} 
\Bigl\langle \frac{|\vec{P}_{h\perp}|}{M}\,  \sin{(\phi_h - \phi_S)} \Bigr\rangle_{UT}^{\pi^{\pm,0}} (x,z) \approx
\frac{3}{2}\,z\,\frac{f_{1T}^{\perp (1)\sca}(x)}{f_1^u(x)}
+ \frac{1}{2}\,z\,\frac{f_{1T}^{\perp (1)\ps}(x)}{f_1^u(x)}\,
\biggl(1 +
\frac{1}{2}\,\frac{D_1^{d(\pi^{\pm,0})}(z)}{D_1^{u(\pi^{\pm,0})}(z)}\biggr).
\end{equation} 

\section{Conclusions}

We calculated the Sivers function in a spectator model of the nucleon
with scalar and
axial-vector diquarks. The final state interaction necessary to generate T-odd
distribution functions was provided by gluon rescattering between the
struck quark and the diquark.

The inclusion of axial-vector diquarks allowed us to
calculate 
the Sivers function for the $d$ quarks. The function turns out to have the
opposite sign compared to the $u$ quarks and to be much smaller in
size. The $u$ quark Sivers function is substantially reduced by the
axial-vector contribution.
Although the reliability of the model is very limited, we think that our
results on the relative behavior of $u$ and $d$ quarks could be qualitatively
relevant. 

Using the results of our model, we estimated some single 
spin asymmetries containing the Sivers function. These asymmetries are at
present being measured by the HERMES and COMPASS collaborations. 
We noticed that the inclusion of axial-vector diquarks can make drastic 
changes in the asymmetries as compared to the spectator model with scalar
diquarks only, particularly at low $x$.
We were able for the first time to estimate the Sivers single spin asymmetry
in $\pi^-$ and $\pi^0$ production. We observed that the $\pi^+$ and $\pi^-$ asymmetries
are not very different, due to the dominance of the $u$ quark
contribution in both cases. The  $\pi^0$ asymmetries (which we did not show) lie  between the $\pi^+$ and $\pi^-$ estimates.

The present approach does not take into account sea quarks. Unfortunately, at
the moment we have no indication about the size and sign of the
sea-quark Sivers function.


\begin{acknowledgments}
Discussion with A.~Metz, D.~S.~Hwang and M.~Oettel, M.~Stratmann 
are gratefully aknowledged.
The work of A.~B.\ has been
supported by the TMR network HPRN-CT-2000-00130 and the BMBF, 
the work of J.~Y.\ by
the Alexander von Humboldt Foundation and by the
Foundation for University Key Teacher of the
Ministry of Education (China).
 
\end{acknowledgments}


\bibliographystyle{apsrev}
\bibliography{mybiblio}

\end{document}